\magnification=\magstep1
\baselineskip=17pt
\hfuzz=6pt

$ $

\centerline{\bf Black Holes, Demons and the Loss of Coherence:}

\bigskip

\centerline{How complex systems get information, and what they do with it.}

\vskip 1cm

\centerline{Seth Lloyd}

\vskip 1cm

\centerline{Ph.D. Thesis}

\centerline{Theoretical Physics}

\centerline{The Rockefeller University}

\centerline{April 1, 1988}

\vskip 1in

\noindent{\it Chapter 3: Pure State Quantum Statistical Mechanics and Black Holes}

If one flips a coin twenty times, and gets heads nine times and tails eleven, one ascribes the variation in the results to differences in how hard one flipped the coin, how it hit the ground, etc. If one prepares twenty electrons in the state spin 
$x$ 
up, then makes a measurement of spin z on each electron and gets spin
$z$
up nine times and spin
$z$
down eleven, then (if one does not believe hidden variable theories) one ascribes the variation to the statistical properties of pure quantum mechanical states.

In quantum statistical mechanics, the inherently statistical nature of quantum mechanical pure states adds an additional element of chance to the already chancy results of measurements made on systems with many degrees of freedom, not all of which have been fixed by experiment. The sort of probabilities that come out of quantum mechanical pure states differ significantly from those that come out of classical probability distributions. In this chapter we show, however, that the pure states of quantum mechanical systems with many degrees of freedom reproduce the statistics of the normal statistical mechanical ensembles such as the microcanonical and canonical ensemble to a high degree of accuracy: quantum mechanical systems with many degrees of freedom in pure states behave like statistical mixtures, with respect to most measurements. In the thermodynamic limit, as the number of degrees of freedom of the
system in question goes to infinity, the difference between the statistics implied by the pure states of the system and the statistical mechanical probabilities becomes impossible to detect.

Suppose, for example, that one has a gas composed of $n$ particles, confined to a box with sides of length $L$. 
Classically, if one fixes the state of the gas by fixing the position and momentum of each particle, then the results of any measurement that one makes on the gas are fixed as well. Quantum mechanically, if one puts the gas as a whole in a pure state by putting each of the particles in a pure state that is an eigenstate of the energy and momentum for a particle in a box of the given size, then only the results of measurements that correspond to operators that commute with the particles' momenta are fixed: if one makes a measurement on the position of a particle with energy and momentum high compared with
$\hbar/L$,
the particle can turn up in any part of the box with equal probability. This uniform distribution for the position of the particle in the box arises from the form of the pure state of the gas as a whole, but it is exactly the same as the distribution predicted by the normal statistical mechanical ensemble for the positions of the particles of the gas with total energy and momentum fixed.

If a quantum mechanical system is in a pure state, then only measurements that correspond to operators of which that state is an eigenstate will give the same result each time. Other measurements will give a statistical distribution of results when performed on an ensemble of systems all prepared in that state. We will prove that for a given
measurement, if many copies of a system with many degrees of freedom have all been prepared in the same pure state chosen at random from a subspace of Hilbert space (such as the subspace
$H_{E,E+dE}$
composed of all states with energy between 
$E$ and $E+dE$),
the results of the measurement made on these systems in this state lie in a statistical distribution whose mean, standard deviation, and higher moments differ from the mean, standard deviation, and higher moments predicted for the results of the measurement by a conventional uniform distribution (the microcanonical
ensemble, for the subspace
$H_{E,E+dE}$)
over all states in that subspace by a factor
of $1/\sqrt n$,
on average, where
$n$
is the dimension of the subspace. For a given measurement, most pure states of a system with many degrees
of freedom give statistical distributions of results that differ by only a small amount from those predicted by the ensemble average. Note that this result has no classical analogue: many copies of a classical system all prepared in the same state all give the same result for any measurement made upon them -- they imply no distribution whatsoever.

As a corollary, we prove that for many copies of a quantum mechanical system with many degrees of freedom prepared in a given state (for example, a state with energy between
$E$ and $E+dE$),
the results of most measurements restricted to a given subspace
$H_{E,E+dE}$
fall in distribution that have expectation values, standard deviations, and
higher moments that differ by only a small amount
($1/{\rm dim}( H_{E,E+dE} )^{1/2}$)
from the expectation values, standard deviations, and higher moments predicted by the ensemble average over that subspace (the microcanonical
ensemble for states with energy between
$E$ and $E+dE$.)
For a system with many degrees of freedom, the quantum mechanical probabilities inherent in the actual state of the system are likely to mimic the normal statistical mechanical probabilities. In the thermodynamic limit,
$n\rightarrow \infty$,
the correspondence is exact.

Another way of phrasing this result is that systems with many degrees of freedom in pure states behave as if they were in statistical mixtures with regard to most measurements We apply this result to the
foundations of statistical mechanics, to black holes, and in chapter 4
to the quantum measurement problem.

Our results follow directly from

\bigskip\noindent
Theorem 1:$^{1-2}$

Let $H \equiv C^n$ be a subspace of the Hilbert space for a
quantum mechanical system $A$.  Let $F$ be an Hermitian operator
corresponding to the measurement of some quantity on $A$.  Then
$$  \big( ~  \langle~ (~ \langle \psi | F | \psi\rangle 
- {\rm tr} F/n~ )^2~ \rangle_{|\psi\rangle \in H}~ \big)^{1/2}   
= {1\over(n+1)^{1/2}} 
\big(~  {\rm tr} F^2/  n -  ({\rm tr} F/n)^2  \big)^{1/2}
.\eqno(1)$$
Proof in appendix.

In the case that
$H$
represents the Hilbert space of states compatible with the results of macroscopic measurements that have been performed on the system, we may paraphrase this theorem as follows: If the dimension of the Hilbert space of states compatible with our macroscopic knowledge of the system is large, then the amount by which the expectation value
of an operator 
$F$
on a typical state 
$|\psi\rangle$
differs from its average expectation value over all compatible states is likely to be small. Indeed, since
$$
\big(  {\rm tr} F^2 / n -  ({\rm tr} F /n)^2   \big)^{1/2}
\leq {\rm max} |f_i|, \eqno(2)$$
where the $f_i$
are the possible results of the measurement, the theorem above implies that the amount by which the expectation value of an operator on a typical state differs from its expectation value over all
states is likely to be less than 
$1/\sqrt n$ times the maximum magnitude of the result.

For example, if a number of systems identical to
$A$ have all been prepared in the same pure state 
$|\psi\rangle \in H_{E,E+dE}$,
the theorem above tells us that if 
$n = {\rm dim} H_{E,E+dE}$
is large, 
the distribution of results for a measurement corresponding to a Hermitian operator
$F$
are likely to have the same expectation value as that implied for the measurement by the microcanonical ensemble. Of course, if
$|\psi\rangle$
is an eigenstate of $F$,
or a superposition of states dominated by a single eigenstate of 
$F$,
 the distribution of the different results,
$f_i$, 
of the measurement implied by
$|\psi\rangle$
differs markedly from the statistics given by the microcanonical ensemble.
If we
make a measurement corresponding to 
$F$
on a number of systems all of which have been prepared in some eigenstate of
$F$,
we will obtain the same result for each system. The microcanonical ensemble, in contrast, predicts a range of results. The theorem above says that if
$n$
is large, the probability that a state selected at random will be a superposition dominated by a single eigenstate of 
$F$
is small.

The proof of the following corollary follows immediately from the proof of the theorem above:

\bigskip\noindent{\it Corollary:}

Given a state $|\psi\rangle \in H \equiv C^n$, 
and a set of real numbers $\{ f_i \}$,
the average over all  {\it bases} $\{ |e_i\rangle \}$ for $H$ of
$$ (~ \langle \psi | F | \psi\rangle
- {\rm tr} F/n~ )^2$$
is equal to 
$$ {1\over(n+1)^{1/2}}
\big(  {\rm tr} F^2/n -  ({\rm tr} F/n)^2  \big)^{1/2},
\eqno(3)$$
where $F=\sum_i f_i |e_i\rangle\langle e_i|$.

This corollary has the consequence that almost all measurements that we can make on a number of systems all prepared in a particular state whose energy lies between 
$E$ and $E+dE$
will
 have expectation values very close to those predicted by the microcanonical ensemble. Even if some clever fellow has prepared our system in a very particular state, most of the measurements that we perform on the system will have results that follow a microcanonical distribution. To put the same point in a different way: if we do not know in what pure state a system has been prepared, the chances of our making a measurement that has that state as one of its eigenstates are slim.

\bigskip\noindent
We have the following results: 

\item{(1)} The expectation value of a particular operator over most quantum states of a system with n degrees of freedom, and with energy in the interval 
$[E,E+dE]$,
is likely to be
the same as the operator's expectation value over the microcanonical ensemble to within a factor of 
$1/\sqrt n$.

\item{(2)} The expectation values of most operators over a particular quantum state with energy in the interval
$[E,E+dE]$
are likely to be equal to those operators' expectation values over the microcanonical ensemble to within a factor of $1/\sqrt n$.

We can go further. Not only do the pure quantum states of a system with many degrees of freedom give expectation values for measurements that are very close to the expectation values predicted by the microcanonical ensemble, these pure states also give standard deviations from those expectation values that are very close to the standard deviations predicted by the microcanonical ensemble.

For a particular measurement, 
$F$,
the microcanonical ensemble over
$H_{E,E+dE}$
predicts a variance from the mean value of
$F$ of
$ {\rm tr} F^2/n - ( {\rm tr} F/n )^2$,
where the traces are taken over
$H_{E,E+dE}$
For an
ensemble of systems prepared in a pure state 
$|\psi\rangle$,
the variance of the results of a measurement of 
$F$ from $(1/n) {\rm tr} F$
is 
$\langle \psi| (F - {\rm tr} F/n)^2 |\psi\rangle$.
Let us look at the average amount by which the variance of
$F$ over $|\psi\rangle$
differs from $F$'s 
variance over the microcanonical ensemble, i.e., let us look at
$$\bigg( 
\langle ~ 
\big( 
~\langle \psi| (F -  {\rm tr} F/n  )^2 |\psi\rangle
- ( {\rm tr} F^2 /  n  - ( {\rm tr} F/n)^2 ~ 
\big)^2 ~ 
\rangle_{|\psi\rangle \in H_{E,E+dE} } 
\bigg)^{1/2}.\eqno(4)$$

By the theorem above, this quantity is equal to
$$ {1\over(n+1)^{1/2} }
\big(~
{\rm tr} B^2/n - ({\rm tr} B/n )^2 ~  
\big)^{1/2}, \eqno(5)$$
where 
$B = (F - {\rm tr} F/n )^2$.

The quantity
$\big(
{\rm tr} B^2/n - ({\rm tr} B/n)^2
\big)^{1/2}$ 
is on the order of the
average magnitude of an eigenvalue of $F^2$,
and the factor of
$1/(n+1)^{1/2}$
out front then insures that as 
$n$
gets large,
the amount by which the variance given by the pure state quantum statistics differs from the variance implied by the microcanonical ensemble tends to become small.

One can apply the same argument to the higher moments of the distribution. Since the average over all
$|\psi\rangle$
of 
$$ \big( \langle \psi| F^m|\psi\rangle - {\rm tr} F^m/n \big)^2
=
{1\over n+1} \big(~
 {\rm tr} F^{2m}/n - ({\rm tr} F^m / n)^2 ~ \big), \eqno(6)$$
as $n$ gets large the moments given by the pure quantum states converge on the microcanonical moments.

For quantum systems with more and more degrees of freedom, not only do the expectation values implied by the pure quantum states of the system tend to mimic more and more exactly the expectation values of statistical mechanics -- the quantum deviations away from those expectation values get closer and closer to the deviations predicted by statistical mechanics, as well. In the thermodynamic limit,
$n\rightarrow\infty$,
the probability that a pure state selected at random from
$H_{E,E+dE}$
gives a distribution for the results of a given measurement that differs from the distribution implied by the microcanonical ensemble, is zero.

We now apply the results derived above to the canonical and grand canonical ensembles.

\bigskip\noindent{\it The Canonical Ensemble}

In this section we show that the exact state for a system in 
contact with a thermostat at temperature $T$ is likely to be a 
mixture that has the same form as the canonical ensemble for the system.

Suppose we have a joint system $AB$. If $A$ and $B$ are weakly interacting, then their Hilbert space can be decomposed into the tensor product space:
$H_{AB} = H_A \otimes H_B$. lowest order in perturbation theory, the subspace
$H^E_{AB}$ of $H_A \otimes H_B$ corresponding to energy $E$ is spanned by a 
basis of vectors of the form $|E_i\rangle_A^k \otimes |E_j\rangle^l_B$,
where $E_i+E_j = E$ and the $|E_i\rangle^k_A$, $k=1$ to
$d_A(E_i)$ ($d_A(E_i)$ is the degeneracy of $E_i$ in $H_A$) span the subspace
of $H_A$ corresponding to energy $E_i$; similarly for the $|E_j\rangle_B^l$,
$l=1$ to $d_B(E_j)$.

A typical state $|\psi\rangle \in H^E_{AB}$ can be written
$$|\psi\rangle = \sum_i ~ \sum_{k=1}^{d_A(E_i)}
~ \sum_{l=1}^{d_B(E-E_i)} \alpha_{kl}^i |E_i\rangle^k_A 
|E-E_i\rangle^l_B. \eqno(7)$$
The corresponding density matrix is
$$\eqalign{ \rho  =  |\psi\rangle \langle \psi| 
&= \sum_{i,i'}~ \sum_{k, k'=1}^{d_A(E_i), d_A(E_{i'})} 
~  \sum_{l,l'=1}^{d_B(E-E_i), d_B(E-E_{i'})} \cr
&[ ~\alpha_{kl}^i {\bar\alpha}_{k'l'}^{i'} 
|E_i\rangle^k_A{}^{k'}\langle E_{i'}| \otimes 
|E-E_i\rangle^l_B {}^{l'}\langle E - E_{i'}| ~   ],\cr}\eqno(8)$$
and the density matrix for $A$ alone is got by taking the trace
of $\rho$ over the degrees of freedom of $B$:
$$\rho_A = \sum_i ~ \sum_{l=1}^{d_B(E-E_i)} ~
\sum_{k, k'=1}^{d_A(E_i)} 
~\alpha_{kl}^i {\bar\alpha}_{k'l'}^{i'}
|E_i\rangle^k_A{}^{k'}\langle E_{i}|.\eqno(9)$$
Note that $\rho_A$ has no off-diagonal terms between states of
different energy.  In general, for a particular pure state $|\psi\rangle$
with energy $E$, $A$ is in a mixture of states with different
energies $E_i$, each correlated with a state of $B$ with 
energy energy $E-E_i$.

We can write $\rho_A = \sum_i p(E_i) \rho_{E_i}$,
where $\rho_{E_i} \in {\bar H}_A^{E_i} \otimes H_A^{E_i}$
is a density matrix for a system with energy $E_i$, ${\rm tr} ~
\rho_{E_i} = 1$: the $p(E_i)$ are the probabilities 
that a meassurement of energy of $A$ will find the value
$E_i$.  ${\rm Tr} \rho_A = {\rm tr} \rho_{E_i} = 1$ implies
that $\sum_i p(E_i) = 1$.  We can now ask, if we select
$|\psi\rangle \in H^E_{AB}$ at random, what are the most 
probable values for $p(E_i)$, and how much deviation do we
expect from those values.

Applying theorem 1 above, we know that the most likely 
value of $p(E_i)$ is given by its value over the ensemble
average.  For a particular $\rho_{AB}$, 
$p(E_i) = {\rm tr} \rho_{AB} P_A^{E_i} \otimes
p_B^{E-E_i}$, where $P_A^{E_i}$, $P_B^{E-E_i}$ are the 
projection operators onto the $E_i$ eigenspace of $H_A$
and the $E-E_i$ eigenspace of $H_B$.  Over the ensemble,
$p(E_i)$ averages out to $n_i/n$, where
$n_i = d_A(E_i) d_B(E-E_i)$; i.e., over the ensemble, the
probability of finding a certain energy $E_i$
for $A$ is just proportional to the degeneracy of
such states.  Theorem 2 now tells us that $p(E_i)$
deviates from its most likely value by 
$${1/(n+1)^{1/2}} ~\{ ~ (n_i/ n) - (n_i/n)^2 ~\}^{1/2} 
\approx (\sqrt{n_i})/n, \eqno(10)$$
on average.  So we have
$$\rho_A = (1/n) \sum_i d_A(E_i) d_B(E-E_i) ( 1\pm 1/\sqrt{n_i}) 
\rho_{E_i}.\eqno(11)$$
But $d_A(E_i) = e^{S_A(E_i)}$, and $d_B(E-E_i)
= e^{S_B(E-E_i)}$, where $S_A$ and $S_B$ are the
entropies of $A$ and $B$.  If $B$ has many more degrees of
freedom than $A$, then terms with $E_i$ near zero dominate
the sum and we can approximate $S_B(E-E_i) =
S_B(E) - E_i/T$, where $1/T = \partial S_B/\partial E$,
and we have
$$\rho_A = (1/N) \sum_i ~ \{ ~ 
e^{-(E_i - TS_A(E_i))/T}~ (~ 1\pm 1/\sqrt{n_i} ) ~ \} ~ \rho_{E_i},
\eqno(12)$$
where $1/N = e^{S_B(E)}/n$ and ${\rm tr} \rho_{E_i} = 1$.

We see that for a system $A$ in contact with a thermostat
$B$ at temperature $T$, for almost all pure states
$|\psi\rangle \in H^E_{AB}$, the {\it exact state}
of $A$ is likely to give (to within a small fluctuation) the
same density matrix as the canonical ensemble for $A$.  In
the thermodynamic limit, $n_i \rightarrow\infty$, and the probability
that the density matrix for $A$ differs from a thermal
density matrix becomes zero.

\bigskip\noindent
{\it Non-equilibrium pure state statistical mechanics}

The results derived so far imply that if the initial state for a system with many degrees of freedom is chosen at random, the distributions of results for measurements that that state implies do not differ from the predictions of the normal statistical mechanical ensembles in the thermodynamic limit. We now derive a non-equilibrium result. We show that if a system 
$A$ 
has an arbitrary weak interaction with a larger system
$B$, 
then even if $A$ and $B$ start out in a pure state far from equilibrium,
their joint state evolves into one in which the density matrix for
$A$ has a thermal form. Suppose that
$A$ and $B$ are initially noninteracting with total energy $E$,
as in the previous section, and that one perturbs the system by adding a small interaction Hamiltonian to the original Hamiltonian:
$H' = H + H_{int}$. Suppose that the
initial state for $AB$ is $|\chi_0\rangle$, $H|\chi_0\rangle 
= E |\chi_0\rangle$. 
 According to first order degenerate perturbation theory, the eigenvectors of
$H_{int}$ confined to the subspace of energy $E$ are,
to lowest order, arbitrary orthogonal linear combinations of the unperturbed states of the subspace. If we wait an amount of time 
$t\approx\hbar/\Delta E$, where $\Delta E$ 
is the average energy level spacing of $H_{int}$,
the evolution of the system will take $|\chi_0\rangle$
 to a state that is an arbitrary superposition of the original states
$|E_i\rangle_A|E-E_i\rangle_B$ with total energy $E$.
Applying the results of the section on the canonical ensemble above, the density matrix for
$A$ is then
$$ \rho_A 
= (1/N) \sum_i ~ \{ ~e^{-(E_i - TS_A(E_i))/T}~ (~ 1\pm 1/\sqrt{n_i} ) ~ \} ~ \rho_{E_i},
\eqno(13)$$
as above.  Once again $\rho_A$ 
takes a thermal form.	In the thermodynamic limit, an arbitrarily chosen
$H_{int}$ causes $A$ and $B$ to 
to evolve into a state in which the density matrix for $A$
is is exactly that for $A$ at thermal equilibrium.

\bigskip\noindent{\it Grand canonical ensembles}

In the above analysis, the only requirement on $E$ 
was that it be an additively conserved quantity. We can repeat the steps above including other additively conserved quantities. For example, if the total electric charge of
$A$ and $B$ is $Q_{tot}$, 
then we find that the most likely exact state for $A$ is
$$\rho_A = (1/N') \sum_{i,Q} \rho_{E_i Q} ~\{ ~
e^{-( E_i - \Phi Q - TS_A(E_i,Q))/T} (~ 1\pm 1/\sqrt{n_iQ} ~) ~\},
\eqno(14).$$
where
$\rho_{E_iQ} \in {\bar H}_A^{E_iQ} \otimes  H_A^{E_iQ}$
has trace $1$, $n_{iQ} = d_A(E_i,Q) d_B(E-E_i, Q_{tot} - Q)$,
and 
$\Phi = - T(\partial S_B/\partial Q)|_{Q=Q_{tot}}$
is the electric potential of $B$.

The exact state of a system $A$ in contact with reservoirs of heat, charge, particle \#,
etc.,
is likely to give to within a small fluctuation the same density matrix as the various grand canonical ensembles for $A$. 
In the limit that the size of the reservoirs goes to infinity, the probability that the density matrix for $A$
takes on an exact grand canonical form goes to one.

\bigskip\noindent{\it Applications of pure state quantum
statistical mechanics}

We apply the results derived above in three areas: 
1) the interpretation of quantum statistical mechanics, 
2) horizon radiation from black holes and cosmological particle production, 
and 3) the quantum measurement problem (chapter 4).

\bigskip\noindent{\it The interpretation of probability in
statistical mechanics}

Historically, there are two approaches to interpreting the probability distribution function in classical statistical mechanics. In the ensemble approach, the probabilities that describe a complex system are taken to refer to an imaginary ensemble of systems identical to the one of interest, whose states are spread out uniformly over the set of microscopic states consistent with our macroscopic information. In the ergodic approach, the probability distribution is regarded as
giving the relative frequencies of occurrence of the microscopic states of the actual system of interest, averaged over time. For example, if all we know about a system is that it has energy between
$E$ and $E+dE$,
then we describe the system by the microcanonical ensemble, characterized by a probability distribution that assigns equal probability to all states with energy in the interval
$[E,E+dE]$. 
In the ensemble approach, this distribution is taken to refer to an imaginary ensemble of identical systems whose states are spread out evenly over the hypersurface in phase space made up of points representing states with energy between
$E$ and $E+dE$.
In the ergodic approach the microcanonical ensemble is held to describe the actual system of interest in accordance with the ergodic hypothesis -- the trajectory of the representative point of the system is hypothesized to fill up the hypersurface of energy
$E$ 
uniformly, spending an equal amount of time in equal volumes of phase space.
When Boltzmann proposed the first version of the ergodic hypothesis
in 1871,$^3$
one of his goals was to provide a purely mechanical description of the approach to equilibrium, independent of statistical
considerations. In introducing his famous H-theorem in 1872,$^4$
he claimed to have proved the second law of thermodynamics on purely mechanical grounds: any initial distribution of kinetic energy, Boltzmann asserted, would eventually approach Maxwell's distribution. It was not until several years later, prodded by Loschmidt, that he acknowledged the statistical nature of his proof, inherent in the assumption of molecular chaos.

The results on the statistics of pure quantum states of systems with many degrees of freedom derived above suggest a purely mechanical interpretation of results normally derived statistically by taking ensemble averages -- or rather, a purely
{\it quantum} 
mechanical interpretation. If we prepare a system with many degrees of freedom in
{\it any}
state with energy between $E$ and $E+dE$,
then the great majority of measurements performed on that system will give results that follow closely the statistical predictions of the microcanonical ensemble. The statistical nature of the outcome of such measurements arises not because we do not know what state the system is in (we prepared it in a pure state), nor because we are making measurements over a long period of time (we can make the measurements over as short a period of time as we like as long as we do not conflict with the uncertainty principle), but because
quantum mechanics is inherently statistical. For a system with many degrees of freedom, the quantum mechanical statistics inherent in the pure states of the system converge on the statistics implied by the microcanonical ensemble.

\vfill\eject
\bigskip\noindent{\it Horizon radiation and cosmological particle
production}

\smallskip\noindent{\it i) Horizon radiation}

It is well established$^{5-9}$ 
that a black hole of mass $M$
emits radiation with a thermal spectrum at temperature 
$T = (8 \pi M)^{-1}$, where we have set $G=h=c=1$.
Hawking uses this result to argue that a black hole
behaves like a black body (albeit a strange black body, with negative specific heat) with entropy 
$S = 4\pi M^2$.
 In the normal derivation of black-hole radiance, however, one treats the hole as a feature of the background spacetime -- a geometric, not a thermodynamic object. The thermal form of the radiation arises from the periodicity of the Schwarzschild metric in imaginary time, not from overtly statistical
considerations.  T.D. Lee$^9$ has pointed out that while black-body radiation is incoherent, the radiation coming from the horizon of a hole is fundamentally coherent over spacetime as a whole; Lee argues that the coherence of the horizon radiation implies black holes are not black bodies.

Using methods of field theory in curved spacetime$^{8-9}$
one can show that in the presence of a black hole of mass M the incoming vacuum state
$|0\rangle_{in}$ evolves into a state of the form
$$|0\rangle_{out} 
= (1/N) \sum_i ~ \sum_{k=1}^{d(E_i)} ~
e^{-4\pi ME_i}
|-E_k\rangle^k_{inside} |E_i\rangle^k_{outside} \eqno(15)$$
where $|E_i\rangle^k_{outside}$ are states with total energy $E_i$,
outside the hole
(the energy is determined by an observer in the asymptotically flat area of spacetime; 
$d(E_i) = $ the degeneracy of $E_i$), and the
$|-E_k\rangle^k_{inside}$ are states with
total energy $-E_i$ inside the hole.
The total energy and charge of this state are still zero, but the state of the fields outside the hole is a thermal mixture at temperature 
$T=(8 \pi M)^{-1}$.

To see the thermal form of the radiation, note that the density matrix for the fields over the whole of spacetime is
$$\rho_{total}
= {1\over|N|^2}
\sum_{i,i'} \sum_{k,k'=1}^{d(E_i)} e^{-4\pi M (E_i+ E_{i'})}
{|-E_k\rangle^k}_{inside}{}^{k'}\langle -E_{i'}|
\otimes {|E_i\rangle^k}_{outside} {}^{k'}\langle E_{i'}|.\eqno(16)$$
A measurement made outside the horizon corresponds to an operator of the form
$I_{inside}\otimes A_{outside}$, where $I_{inside}$
is the identity operator on the Hilbert space of states inside the hole, and
$A_{outside}$
is an Hermitian operator on the space of states outside the hole. We have
$$\langle I_{inside}\otimes A_{outside}\rangle
= {\rm tr} \rho_{total} I_{inside}\otimes A_{outside}
= {\rm tr} \rho_{outside} A_{outside},\eqno(17)$$
where
$\rho_{outside} = $ the 
the trace over the internal degrees of freedom of the hole of
$\rho_{total}$
$$\eqalign{ =& (1/|N|^2) \sum_{i,i',k,k'} e^{-4\pi M(E_i + E_{i'})}
{|E_i\rangle^k}_{outside} {}^{k'}\langle E_{i'}| 
\delta_{ii'} \delta_{kk'}\cr
=& (1/|N|^2) \sum_{i,k} e^{-8\pi M E_i}
{|E_i\rangle^k}_{outside} {}^{k}\langle E_{i}|.\cr}\eqno(18)$$
Note that even though the state of the fields over the whole of spacetime is pure, the fields outside the hole are in a mixture. In fact, the form of
$\rho_{outside}$
is exactly that of a thermal mixture at temperature 
$T=(8\pi M)^{-1}$. 
The hole radiates with a black-body spectrum.

T.D. Lee has argued$^9$ that the coherence of the quantum fields over the whole of spacetime implies that black holes should not be regarded as black bodies, the radiation 
from which is incoherent. Our results from the previous section 
imply the opposite: {\it
the form of the coherent fields in the presence of a black hole is exactly what we expect of the coherent fields in contact with a black body in the thermodynamic limit.
}
The results of the previous sections on the canonical ensemble imply
that the pure state of the quantum fields interacting with a black body of temperature
$T$
is very likely to take the form of a thermal mixture at temperature 
$T$.
The results on non-equilibrium pure state statistical mechanics derived above imply that if the quantum fields around the hole are not initially in a thermal form, an arbitrary interaction between the field modes and the degrees of freedom of the gravitational field will induce the state of the fields to attain a thermal form, at least locally (since black holes have negative specific heat, the fields can not be in global stable equilibrium with the hole). If a black hole is a black body with temperature 
$T= (8\pi M)^{-1}$,
then the factor by which the form of the coherent state of the fields interacting with the hole
deviates from the form of a thermal mixture is on the order of
$e^{-2\pi M^2} = e^{-(2\pi \times 10^{76})}$ for $M= M_{\odot}$.  
Black holes may not be black bodies, but if they aren't, it is very difficult to tell the difference.

\bigskip\noindent{\it
ii) Increase of mutual information and cosmological particle production}

The theorems on Hilbert space proved above allow us to carry further
the arguments of Hu and Kandrup,$^{10}$ 
who treat the problem of cosmological particle generation and entropy production in terms of increase of the mutual information shared between the oscillators that represent the modes of quantum fields in Fock space. Kandrup and Hu show that increase in the mutual information between the oscillators is closely linked to cosmological partical production. (As we have seen, a black hole creates particles with induced entropy
$-{\rm tr} \rho_{outside} \ln \rho_{outside}$
outside
the hole, even though the fields over all of spacetime are in a pure state.) 
They point out that the mutual information is never negative, and if the
mutual information is initially zero, any interaction will cause it to increase, at least initially.

We use the non-equilibrium results derived above to show that the mutual information shared between the particle modes of the quantum fields in the universe is very likely to increase to its maximum value.

The modes of the quantum fields can be represented as a collection of harmonic oscillators: we first examine the case of n operators without interactions, then introduce interactions and take the thermodynamic limit
$n\rightarrow\infty$.  $n$ 
noninteracting harmonic oscillators are described by a Hamiltonian
$H= H_1 + H_2 + \ldots + H_n$, where
$H_l = \sum_{i=0}^\infty i\hbar \omega_l |i\rangle_l\langle i|$, 
where $|i\rangle_l$ is the $i$-th excited state of the $l$-th 
oscillator, $\omega_l$ is 
is its fundamental frequency, and we have set the zero point energy to zero. Suppose that the oscillators start out in the state 
$|\chi_0\rangle = |i\rangle_1|j\rangle_2 \ldots |k\rangle_n$
with total energy $E = \hbar(i\omega_1 + j\omega_2 + \ldots + k\omega_n)$.
If there is no interaction between the oscillators, then they will stay in this state forever. Suppose now that one perturbs the system by adding a small interaction Hamiltonian to the original Hamiltonian:
$H' = H + H_{int}$.
According to first order perturbation theory, the elgenvectors of 
$H_{int}$ 
confined to the subspace of energy
$E$
are, to lowest order, arbitrary orthogonal linear combinations of the unperturbed states of the subspace. If we wait an amount of time
$t\approx\hbar/{\Delta E}$, where $\Delta E$ is the average energy level spacing
of $H_{int}$, 
the evolution of the system will take $|\chi_0\rangle$ 
to a state that is an arbitrary superposition of the original states 
$|i\rangle|j\rangle \ldots |k\rangle$
with total energy $E$.
Applying the results of the section on the canonical ensemble above, the density matrix for the first oscillator is then
$$\rho_1 = (1/N) \sum_{i=0}^\infty
|i\rangle_l\langle i| d_{2\ldots n}(E-\hbar\omega_l i) 
~ ( ~ 1\pm 1/(d(E-\hbar\omega_l i)^{1/2} ~) \eqno(19)$$
where $d_{2\ldots n}(E-\hbar\omega_l i)$ 
is the dimension of the space of states of the oscillators 
$2\ldots n$, with total energy $E-\hbar\omega_l i$, $N$ is a 
normalization constant, and the $\pm 1/(\ldots)^{1/2}$ 
expresses the uncertainty in the result due to the lack of knowledge of the exact form of
$H_{int}$.  
If $E$ is large and $n>>1$, we can write
$$  d_{2\ldots n}(E-\hbar\omega_l i)
= e^{S_{2\ldots n}(E-\hbar\omega_l i)}
= e^{S_{2\ldots n}(E) - (\partial S_{2\ldots n}/\partial E) \hbar \omega_l i}
.\eqno(20)$$
Taking the thermodynamic limit $n\rightarrow\infty$, we obtain
$$\rho_1 = (1/N') \sum_{i=0}^\infty
|i\rangle_l\langle i| e^{-\hbar \omega_l i/T}, \eqno(21)$$
where
$1/T = \partial S_{2\ldots n}/\partial E$.
The expressions for $\rho_2, \ldots, \rho_n$ are identical,
with appropriate re-indexing.

 In the thermodynamic limit, an arbitrarily chosen $H_{int}$
causes the oscillators as a group to evolve into a state in which the density matrix for each of the oscillators is exactly that for the oscillators at thermal equilibrium; the evolution drives both the mutual information and the coarse-grained entropy up to their maximum values, equal to the equilibrium thermodynamic entropy.

Our results imply that correlations exhibited by interacting quantum systems in a pure state are not only likely to increase from an initial
zero value, but that they are likely to tend to continue rising and
attain an equilibrium value given by the normal statistical mechanical entropy for the systems.

\vskip 1cm
\noindent{\it Appendix: proof of theorem 1}

We want to prove the following: Given an Hermitian operator $F$ on
$H \equiv C^n$, then the average over all
$|\psi\rangle \in H$ of $ ( \langle \psi| F |\psi\rangle
-(1/n) {\rm tr} F )^2$ 
$$= 1/(n+1) \big( ({\rm tr} F^2)/n 
- ({\rm tr} F)^2/n^2 \big).\eqno(A1)$$

\smallskip\noindent Proof:

Pick a basis $\{ |e_i\rangle \}$ for $H$ in which $F$ is diagonal:
$F=\sum_i f_i |r_i\rangle \langle e_i|$.  The average over all
$|\psi\rangle = \sum_i \alpha_i |e_i\rangle \in H$
of $ ( \langle \psi| F |\psi\rangle
-(1/n) {\rm tr} F )^2$ is equal to the average over
$(\alpha_1, \ldots,\alpha_n) \in$ the unit sphere of $C^n$ of
$$ \big( \sum_{i=1}^n f_i (|\alpha_i|^2 - 1/n) \big)^2, \eqno(A2)$$
which is equal to the average over the unit sphere of
$$\sum_{i,j=1}^n f_i f_j \{ ~ |\alpha_i|^2 |\alpha_j|^2
- (1/n)( |\alpha_i|^2 + |\alpha_j|^2 ) + 1/n^2 ~ \} .\eqno(A3)$$
But the average over the unit sphere of $|\alpha_i|^2$ is $1/n$, the average
of $|\alpha_i|^2 |\alpha_j|^2$ is $2/n(n+1)$ if $i=j$ and is $1/n(n+1)$
if $i\neq j$.  The average over all
$|\psi\rangle \in H$ of 
$ ( \langle \psi| F |\psi\rangle
-(1/n) {\rm tr} F )^2$ is equal to
$$\eqalign{
\sum_{i,j=1}^n f_i f_j & ~(~ 2\delta_{ij}/n(n+1)
+ (1-\delta_{ij})/n(n+1) - 2/n^2 + 1/n^2 ~ )~ \cr
&= 1/(n+1) ~(~ ({\rm tr} F^2)/n - ( {\rm tr} F)^2/n^2 ~). \cr
}\eqno(A4)$$
Taking the square root of both sides completes the proof.

\vskip 1cm
\noindent{\it References}

\bigskip

\noindent 1. This theorem is implied by the proof of the first part of 
Jancel's `first statistical ergodic theorem,' in R. Jancel, 
{\it Foundations of Classical and Quantum Statistical Mechanics,}
 Oxford (1969). Our theorem is not an ergodic theorem, however: 
it applies to the state of a system at a given time.

\smallskip\noindent
2. P. Bocchieri, G.M. Prosperi, in  {\it Statistical Mechanics:
Foundations and Applications,} ed. T.A. Bak,
W.A. Benjamin, (1967).

\smallskip\noindent
3. L. Boltzmann, {\it Wiener Berichte}  {\bf 63},
391 (1871). 

\smallskip\noindent
4. L. Boltzmann, {\it Wiener Berichte} {\bf 66},
275
(1872).

\smallskip\noindent
5. S.W. Hawking,  {\it Nature} {\bf 248}, 30 (1974).

\smallskip\noindent
6. S.W. Hawking,  {\it Commun. Math. Phys.} {\bf 43}, 199 (1975).

\smallskip\noindent
7. J.B. Hartle and S.W. Hawking, {\it Phys. Rev. D} {\bf 13}, 870 (1976).

\smallskip\noindent
8. W.G. Unruh,	 {\it Phys. Rev. D} {\bf 14},
870
(1976).

\smallskip\noindent
9. T.D.	Lee, {\it Nucl. Phys. B} {\bf 264},
437
(1985).

\smallskip\noindent
10. B.L. Hu, H.E. Kandrup, {\it Phys. Rev. D} {\bf 35}, 1776 (1987).

\vfill\eject\end